**Smooth coherent Kerr frequency combs generation with broadly tunable pump by higher order mode suppression**

S.-W. Huang[1*+], H. Liu[1+], J. Yang[1], M. Yu[2], D.-L. Kwong[2], and C. W. Wong[1*]

[1]Mesoscopic Optics and Quantum Electronics Laboratory, University of California Los Angeles, CA, USA

[2]Institute of Microelectronics, Singapore, Singapore

[*]swhuang@seas.ucla.edu, cheewei.wong@ucla.edu

[+]these authors contributed equally to this work

**High-$Q$ microresonator has been suggested a promising platform for optical frequency comb generation, via dissipative soliton formation. To achieve a higher $Q$ and obtain the necessary anomalous dispersion, $Si_3N_4$ microresonators made of multi-mode waveguides were previously implemented. However, coupling between different transverse mode families in the multi-mode waveguides results in periodic disruption of dispersion and quality factor, introducing perturbation to dissipative soliton formation and amplitude modulation to the corresponding spectrum. Careful choice of pump wavelength to avoid the mode crossing region is thus critical in conventional $Si_3N_4$ microresonators. Here, we report a novel design of $Si_3N_4$ microresonator such that single mode operation, high quality factor, and anomalous dispersion are attained simultaneously. The microresonator is consisted of uniform single mode waveguides in the semi-circle region, to eliminate bending induced mode coupling, and adiabatically tapered waveguides in the straight region, to avoid excitation of higher order modes. The intrinsic $Q$ of the microresonator reaches $1.36 \times 10^6$ while the GVD remains to be anomalous at -50 $fs^2$/mm. We demonstrate, with

**this novel microresonator, broadband phase-locked Kerr frequency combs with flat and smooth spectra can be generated by pumping at any resonances in the optical C-band.**

Optical frequency combs, unique light sources that coherently link optical frequencies with microwave electrical signals, have made a broad impact on frequency metrology, optical clockwork, precision navigation, and high speed communication over the past decades [1], [2]. Parametric oscillation in ultrahigh $Q$ microresonators [3], [4], facilitated by the high quality factors and the small mode volumes, is an alternative physical process that offers the opportunity of optical frequency comb generation in compact footprints [5]. The recent demonstration of octave spanning parametric oscillation [6], [7], low-phase noise photonic oscillator [8]–[11], stabilized optical frequency microcomb [12]–[14], and mode-locked femtosecond pulse train [15]–[19] have triggered great excitements. In particular, the observation of dissipative soliton formation and soliton induced Cherenkov radiation [20] offers a reliable route towards self-referenced broadband optical frequency microcomb. Dissipative solitons are localized attractors where the Kerr nonlinearity is compensated by the cavity dispersion and the cavity loss is balanced by the parametric gain [21]. Thus the cavity dispersion and the pump-resonance detuning are two important parameters determining the existence of dissipative solitons in these ultrahigh $Q$ microresonators.

Generation of microresonator-based optical frequency comb, or Kerr frequency comb, has been studied in various material platforms [10], [22]–[25], including $Si_3N_4$ planar waveguide system that is especially suitable for monolithic electronic and feedback integration. For $Si_3N_4$ microresonators, dispersion is typically engineered by design of waveguide geometry. Anomalous dispersion, required for bright dissipative soliton formation, is achieved using multi-mode waveguides in the optical C/L-band wavelength range. Besides, scattering loss in multi-mode waveguides is reduced, leading to higher quality factors and lower comb generation threshold [26]. However, coupling between different transverse mode families in the multi-mode waveguides results in periodic disruption of dispersion and quality factor, introducing additional perturbation to the Kerr frequency comb and dissipative soliton generation dynamics [27]–[31]. Such effect manifests itself as characteristic amplitude modulation in the Kerr frequency comb spectrum or detrimental destabilization of the dissipative cavity soliton, depending on the strength and position of the mode coupling [29], [31]. Careful choice of pump mode to avoid the mode crossing region and increase of cavity's free spectral range (FSR) are thus necessary for dissipative soliton formation in conventional $Si_3N_4$ microresonators.

Here we report a novel design of $Si_3N_4$ microresonator such that single mode operation, high quality factor, and anomalous dispersion are attained simultaneously. No higher order mode is observed throughout the optical C/L-band in the transmission spectrum. A high resolution coherent swept wavelength interferometer (SWI) is implemented to determine the intrinsic quality factor and the group velocity dispersion (GVD) of the microresonator. They are measured at $1.36 \times 10^6$ and -50 $fs^2$/mm, respectively. We demonstrate, with the novel microresonator, phase-locked Kerr frequency combs can be generated by pumping at any resonances in the optical C-band. The spectra spanning more than 20 THz (full width at -20 dB) are smooth without periodic amplitude modulations.

For applications such as high speed communication and astrospectrograph calibration, it is beneficial to have a broadband optical frequency comb with a smooth and flat spectral shape [32], [33]. Such an unstructured spectrum is difficult to attain in a multi-mode microresonator (Figure 1), especially in $Si_3N_4$ microresonators where mode coupling is facilitated by the strong sidewall scattering [26]. Reducing the waveguide cross-section to $1 \times 0.8$ μm$^2$ is a solution to eliminate higher order modes, but it results in larger optical mode overlap with the waveguide boundaries leading to lower quality factor. The propagation loss of a $1\times0.8$ μm$^2$ Si$_3$N$_4$ waveguide is 60% higher than that of a $2\times0.8$ μm$^2$ waveguide, resulting in a 2.5-fold comb generation threshold. The GVD of such single mode waveguide is positive for all optical communication wavelengths (Figure 2a), further inhibiting the Kerr comb generation.

Our strategy to suppress higher order modes while maintaining high quality factor and anomalous dispersion is two-fold (Figure 2b). First, a uniform single mode waveguide ($1\times0.8$ μm$^2$) is included in the semi-circle region where the waveguide is curved, as bending introduces significant mode coupling if a multi-mode waveguide is used. The diameter of the semi-circle is 200 μm. Second, the microresonator is enclosed by adding straight waveguides with adiabatically tapered width to join the two semi-circles. Each straight waveguide is 800 μm long with the width tapered linearly from 1 μm at the two ends to 2 μm at the middle, ensuring selective excitation of the fundamental mode in the otherwise multi-mode waveguide segment. The tapered waveguide accounts for >70% of the cavity length, thus the quality factor is minimally compromised and the average GVD remains to be anomalous in this novel Si$_3$N$_4$ microresonator. The microresonator was fabricated with CMOS-compatible processes: First a 5 μm thick under-cladding oxide was deposited to suppress substrate loss. An 800 nm thick Si$_3$N$_4$ layer was then deposited via low-pressure chemical vapor deposition, patterned by optimized deep-ultraviolet lithography, and etched via optimized reactive ion dry etching. Annealing at a temperature of 1150$^\text{o}$C was then applied to the chip for 3 hours to reduce the waveguide propagation loss. Finally the silicon nitride spiral resonators were over-cladded with a 3 μm thick oxide layer. Characterization of the fabricated microresonator is summarized in Figure 3.

A high resolution coherent SWI was implemented to characterize the cold cavity properties, $Q$ and GVD, of the microresonator [17]. The wavelength of a tunable external cavity diode laser (ECDL) was linearly tuned from 1535 nm to 1625 nm at a scan speed of 60 nm/s. The scan speed was chosen such that the uncertainty of our dispersion measurement, limited by the temperature induced resonance shift, was below 70 kHz/mode. To ensure uniform optical frequency sampling of 7 MHz, 10% of the ECDL output was tapped to an unbalanced Mach-Zehnder interferometer and the photodetector output was fed as the sampling clock of the data acquisition. For absolute wavelength calibration, transmission spectra of the microresonator and a fiber coupled hydrogen cyanide (HCN) gas cell were recorded simultaneously. Fig. 3(a) shows the cold cavity transmission of the microresonator, calibrated with 51 absorption features of the HCN gas cell, and no higher order transverse modes are observed throughout the optical C/L bands in the spectrum. The inset shows the resonance at 1556 nm, which is undercoupled with a loaded $Q$ over 1,000,000 and intrinsic $Q$ about 1,360,000. Fig. 3(b) shows the wavelength dependence of the free spectral range (FSR), measuring a FSR of 64.24 GHz and a mode non-equidistance of 196 kHz at 1556 nm, corresponding to an anomalous GVD of -50 $fs^2$/mm.

A closer look at the transmission spectrum reveals the two resonances around 1558 nm are hybridized modes from polarization coupling (green lines in Fig. 3a) [30]. The strength of the polarization coupling is rather weak and the slight dispersion disruption around 1558 nm is negligible in the Kerr frequency comb formation, evidenced by the smooth spectral shapes shown in Figure 4. Moreover, the 7.5 THz period of the polarization coupling is comparably larger than the 1.5 THz cycle of the higher order mode coupling. Thus the polarization coupling occurs less frequently and introduces fewer perturbations to dissipative soliton formation. We also identify the microresonator temperature, controlled by a thermoelectric cooling device, as a handy parameter to tune the polarization coupling strength. Fig. 3(c) shows the transmission of the same resonance at a different temperature, showing a change of cavity loading from 70% to 40%. It suggests that the polarization coupling effect can be adjusted by stabilizing the microresonator temperature at different levels.

For Kerr frequency comb generation, the microresonator was pumped with an on-chip power up to 1.7 watts. A flat and smooth phase-locked Kerr frequency comb can be generated by pumping at any resonances in the optical C-band, thanks to the absence of quality factor disturbance induced by mode coupling. Fig. 4(a) shows three example spectra pumped at 1551.83 nm, 1556.48nm, and 1561.58 nm. Moreover, the spectral shapes resemble one another and periodic amplitude modulations are not observed, distinctly different from Fig. 1 where the comb spectrum is structured by the mode coupling characteristics. Tuning speed of the pump frequency has been shown to be a critical parameter to drive the Kerr frequency comb into low-phase noise soliton state, circumventing the thermal effect of the microresonator [16]. To find the proper tuning speed, pump power transmission was monitored while pump frequency was scanned, via control of the piezoelectric transducer, across a cavity resonance with tuning speeds varying from 1 nm/s to 100 nm/s. When the tuning speed was higher than 40 nm/s, a characteristic step signature of low-phase noise soliton state was observed. Fig. 4(b) shows the pump power transmission with the pump frequency tuning speed set at 60 nm/s. Coherence of the Kerr frequency comb was characterized by measuring the RF amplitude noise with a scan range much larger than the cavity linewidth [15], [34], [35]. Pump mode was removed with a fiber WDM filter to avoid saturation of the photodetector. Fig. 4(c) shows RF amplitude noise of the Kerr frequency comb at different states, showing the transition from a high-phase noise state (state 1) to the low-phase noise soliton state (state 2). In state 3, the pump was off-resonance and Kerr frequency comb was not generated. At the proper pump-resonance detuning, the RF amplitude noise drops by more than two orders of magnitude (state 1 to state 2) and approaches the background noise of the detector (state 3).

In summary, we report a novel design of $Si_3N_4$ microresonator such that single mode operation, high quality factor, and anomalous dispersion are attained simultaneously. The microresonator is consisted of uniform single mode waveguides in the semi-circle region, to eliminate bending induced mode coupling, and adiabatically tapered waveguides in the straight region, to ensure selective excitation of the fundamental mode. The intrinsic $Q$ of the microresonator is $1.36 \times 10^6$, 1.6 times larger than that of a single mode microresonator with a uniform waveguide cross-section. More importantly, the GVD of the novel microresonator remains to be anomalous at -50 $fs^2$/mm. We demonstrate, with the novel microresonator, phase-locked Kerr frequency combs can be generated by pumping at any resonances in the optical C-band. The spectra spanning more than 20 THz (full width at -20 dB) are smooth without periodic amplitude modulations.

**Acknowledgements:** The authors acknowledge funding support from the DODOS program from DARPA (HR0011-15-2-0014) and the AFOSR Young Investigator Award (FA9550-15-1-0081).

**Author contributions:** S.W.H. and J.Y. designed the microresonator. M.Y. and D.L.K. performed the device nanofabrication. H.L. and S.W.H. conducted the experiment and analyzed the data. S.W.H., H.L., and C.W.W. contributed to writing and revision of the manuscript.

**Additional information:** The authors declare no competing financial interests. Reprints and permission information is available online at http://www.nature.com/reprints/. Correspondence and requests for materials should be addressed to S.W.H. and C.W.W.

\* During preparation of this manuscript, a similar work appeared in [Kordts, A., Pfeiffer, M., Guo, H., Brasch, V., & Kippenberg, T. J. Higher order mode suppression in high-*Q* anomalous dispersion SiN microresonators for temporal dissipative Kerr soliton formation. *arXiv:1511.05381* (2015).]

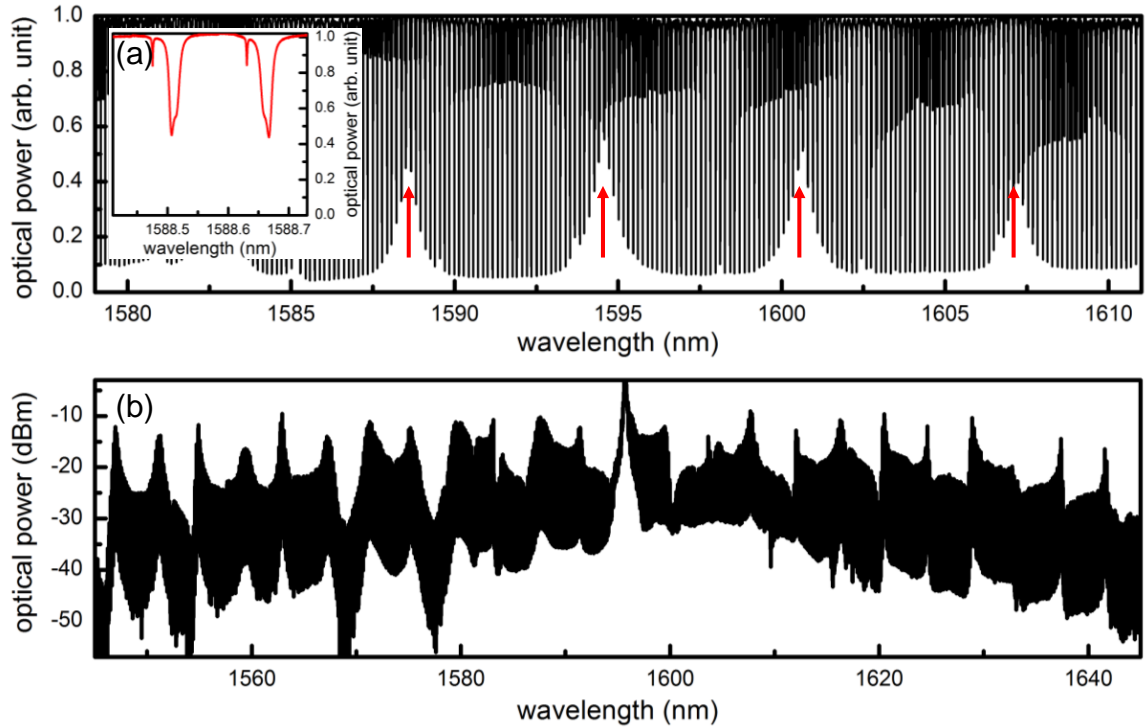

**Figure 1 | Kerr frequency comb from a multi-mode microresonator. a,** Cold cavity transmission of a multi-mode microresonator. The waveguide cross-section is uniform at $2\times0.8$ µm$^2$ along the whole microresonator. Five modal families (3 TE and 2 TM) are identified and mode hybridization between the first two TE modes leads to periodic disruptions in dispersion and quality factor (red arrows). Inset: Zoom-in view of the cavity resonances around 1588.6 nm. **b,** Example Kerr frequency comb spectrum, showing a periodic amplitude modulation in its spectral shape due to the mode hybridization.

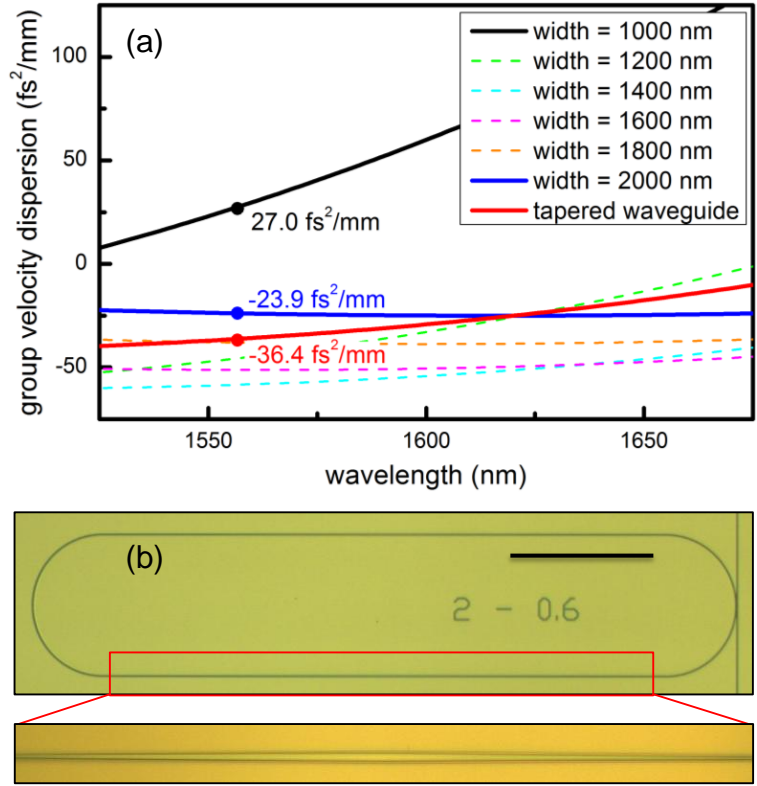

**Figure 2 | Design of a single-mode, high *Q*, and anomalous dispersion microresonator. a,** Group velocity dispersion of the uniform waveguides with different widths and the tapered waveguide calculated with a commercial full-vector finite-element-mode solver (COMSOL Multiphysics), taking into account both the waveguide geometries and the material dispersion. **b,** An optical micrograph of the designed single-mode microresonator. The waveguide in the semi-circle regions has a uniform width of 1 μm, supporting only the fundamental modes. On the other hand, the 800 μm long straight waveguide has a tapered width from 1 μm at the end to 2 μm at the middle of the waveguide. The total cavity length is 2.2 mm. Scale bar: 200 μm.

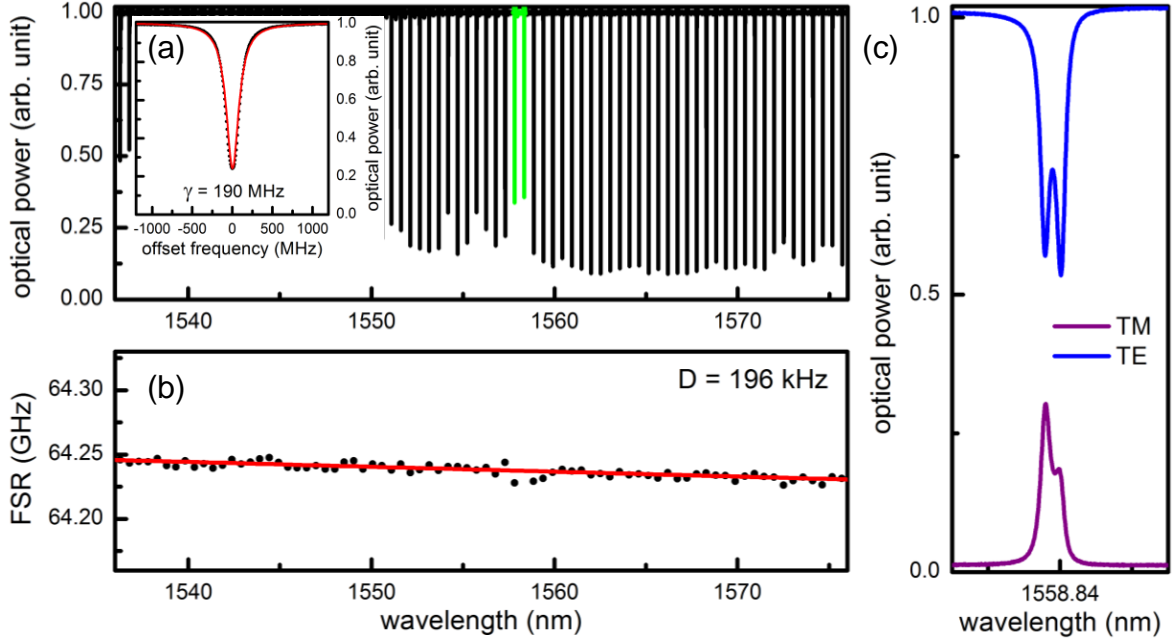

**Figure 3 | Characterization of the single-mode, high *Q*, and anomalous dispersion microresonator. a,** Cold cavity transmission of the designed single-mode microresonator, measured with the high resolution coherent swept wavelength interferometer. Higher order modes are not observed in the microresonator, but the weak TE and TM coupling around 1558 nm results in a 10% reduction in the cavity loading (green lines). Inset: resonance at 1556 nm is undercoupled with a loaded *Q* of 1,000,000. **b,** Wavelength dependence of the free spectral range (FSR), measuring a non-equidistance of the modes, $D = -\frac{\beta_2 c \omega_{FSR}^2}{n}$, of 196 kHz. The extracted group velocity dispersion is anomalous at -50 fs$^2$/mm. The slight dispersion disruption around 1558 nm is negligible in the Kerr frequency comb formation, evidenced by the smooth spectral shapes shown in Figure 4. **c,** Resonances at 1558.84 nm of the microresonator at a different temperature. The input is TE polarized and the output is analyzed by either a TE or a TM polarizer, showing the hybridized modes are superposition of both polarization states.

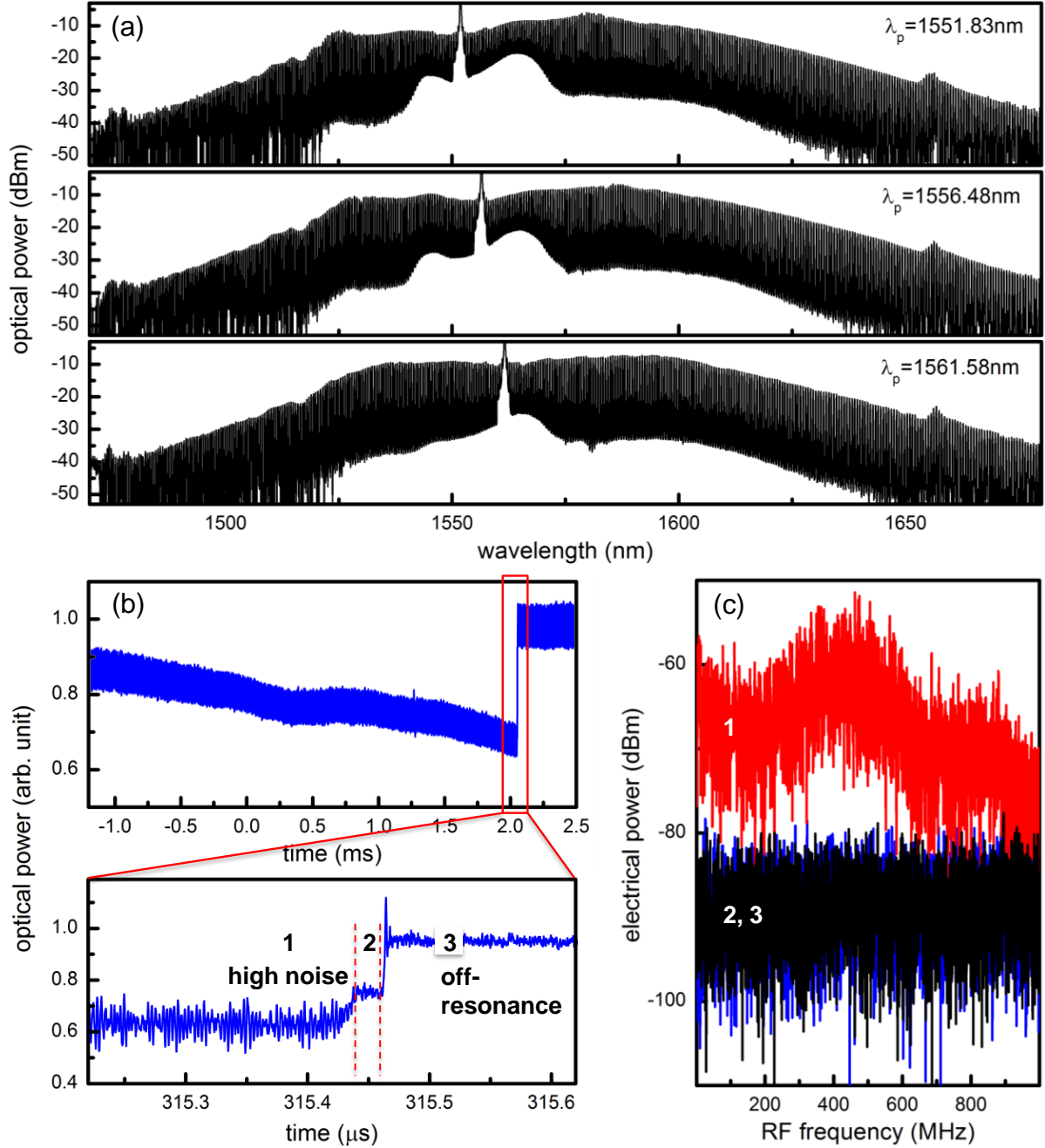

**Figure 4 | Smooth phase-locked Kerr frequency combs pumped at different wavelengths. a,** In the designed single-mode microresonator, phase-locked Kerr frequency combs can be generated by pumping at any resonances in the C-band. Three example spectra are shown here. Moreover, the spectral shapes are smooth without periodic amplitude modulations, distinctly different from Figure 1. **b,** Pump power transmission as the pump wavelength is scanned across a cavity resonance at a speed of 60 nm/s. The observation of the step signature is characteristic of the low-phase noise soliton state (state 2). **c,** RF amplitude noise of the Kerr frequency comb at different states (1: high-phase noise, 2: low-phase noise) along with the detector background (3), showing the transition in and out of the low-phase noise state. The scan range of 1 GHz is more than five times the cavity linewidth.

13